\documentclass[pra,twocolumn,aps,superscriptaddress,showpacs,floatfix]{revtex4-1}
\usepackage{hyperref}
\usepackage{graphicx}
\usepackage{amsmath}
\usepackage{amsfonts}
\usepackage{amssymb}
\usepackage{epsfig}
\usepackage{subfigure}
\usepackage{mathtools}
\usepackage{braket}
\usepackage[usenames,dvipsnames]{color}
\usepackage{setspace}
\usepackage{bm}
\usepackage{ulem}
\usepackage{times}
\usepackage{enumitem}
\usepackage{xcolor}
\hypersetup{
      colorlinks=true,
      citecolor=blue,
      linkcolor=blue,
      urlcolor=blue}
 
\usepackage{filemod}

\def\Tr{\mathrm{Tr}}
\def\be{\begin{equation}}
\def\ee{\end{equation}}
\def\bg{\begin{equation}\begin{gathered}}
\def\eg{\end{gathered}\end{equation}}

\begin{document}
\author{Kuldeep Suthar}
\affiliation{Institute of Theoretical Physics, Jagiellonian University in Krakow,  \L{}ojasiewicza 11, 30-348 Krak\'ow, Poland }
\email{kuldeep.suthar@uj.edu.pl}

\author{Piotr Sierant}
\affiliation{Institute of Theoretical Physics, Jagiellonian University in Krakow, \L{}ojasiewicza 11, 30-348 Krak\'ow, Poland }
\email{piotr.sierant@uj.edu.pl}
\affiliation{ICFO- Institut de Sciences Fotoniques, The Barcelona Institute of Science and Technology, 08860 Castelldefels (Barcelona), Spain}

\author{Jakub Zakrzewski}
\affiliation{Institute of Theoretical Physics, Jagiellonian University in Krakow, \L{}ojasiewicza 11, 30-348 Krak\'ow, Poland }
\affiliation{Mark Kac Complex
Systems Research Center, Jagiellonian University in Krakow, \L{}ojasiewicza 11, 30-348 Krak\'ow,
Poland. }
\email{jakub.zakrzewski@uj.edu.pl}

\date{\today}

\title{Many-body localization with synthetic gauge fields in disordered Hubbard chains}

\begin{abstract}
 We analyze the localization properties of the disordered Hubbard model in the presence of a synthetic magnetic field. An analysis of level 
spacing ratio shows a clear transition from ergodic to many-body localized phase. The transition shifts to larger disorder strengths with 
increasing magnetic flux. Study of dynamics of local correlations and entanglement entropy indicates that  charge excitations remain localized 
whereas spin degree of freedom gets delocalized in the presence of the synthetic flux. This  residual ergodicity is enhanced by the presence 
of the magnetic field with dynamical observables suggesting incomplete localization at large disorder strengths. Furthermore, we examine the 
effect of quantum statistics on the local correlations and show that the long-time spin oscillations of a hard-core boson system are 
destroyed as opposed to the fermionic case.
\end{abstract}

\maketitle

\section{Introduction}
 The phenomenon of many-body localization (MBL) has attracted a significant interest in condensed matter physics over past several years, 
both theoretically~\cite{fleishman_80,alet_18,parameswaran_18,abanin_19} and 
experimentally~\cite{schreiber_15,bordia_16,bordia_16a,smith_16,xu_18}. The MBL is an extension of Anderson localization (which describes 
localization of single-particle eigenstates in the presence of disorder potential) to highly excited eigenstates of interacting many-body 
systems. The characteristic properties of MBL phase are Poisson {eigenvalue} statistics~\cite{oganesyan_07,luitz_15,serbyn_16,vasseur_16,
maksymov_19,sierant_19, Sierant19d}, an absence of thermalization~\cite{pal_10,serbyn_13,luitz_16,sierant_17}, a vanishing 
transport~\cite{agarwal_15,bar_15,kozarzewski_16}, and the logarithmic spreading of the entanglement 
entropy~\cite{bardarson_12,Serbyn13a,mondaini_15}. Starting from early works, many aspects of the topic have been examined to date. The 
ultracold atomic systems are ideal platform to explore the localization phenomena due to their ability to tune dimensionality, nature of 
applied disorder, atomic interactions, lattice geometry and synthetic gauge fields. The existence of MBL phase has been confirmed in recent 
quantum gas experiments using quantum simulators such as optical lattices~\cite{schreiber_15,bordia_16,choi_16,luschen_17} and trapped 
ions~\cite{smith_16}. 

 The nonergodic behaviour of disorder Hubbard chain at strong disorder and temporal evolution of its correlation functions have been examined 
theoretically in~\cite{mondaini_15,prelovsek_16a}. The SU(2) symmetry of the model limits {the full} MBL as the charge degrees of freedom are 
localized but spins remain delocalized and reveal subdiffusive dynamics~\cite{prelovsek_16,kozarzewski_18,sroda_19}. The partial MBL is due 
to the separation of time scales between charge and spin sectors dynamics in the presence of {spin-independent} 
disorder~\cite{zakrzewski_18}. It was {also} suggested that transport {properties} strongly depend on the fraction  of singly occupied sites in the initial state~\cite{protopopov_19}. 
{Still}, the breaking of SU(2) spin symmetry recovers the full MBL phase~\cite{lemut_17,sroda_19,Leipner-Johns19}. 
{The experimental study of a related cold atomic system \cite{schreiber_15,luschen_17} addressed mainly the charge sector observing signatures
of MBL for available experimentally times of about 100 characteristic tunneling times in the optical lattice.} 

{Most of the disorder related studies consider time reversal invariant (TRI) Hamiltonians. Only very recently experimental studies in cold atomic
systems utilizing synthetic gauge fields addresses aspects of single particle localization. The work ~\cite{hainaut_18} studied the significance of symmetry on the localization 
and transport properties in the presence of synthetic gauge field for periodically driven (Floquet) system while \cite{an_18} exploited a lattice in the momentum space
with laser induced couplings to create zigzag chain with the synthetic flux  dependent mobility edge. Single particle localization with synthetic random flux
in a lattice 
has been discussed in \cite{cheng_16,major_17}. Cheng and Mondaini \cite{cheng_16} studied also the model with interaction observing that the presence of random synthetic fluxes 
delocalized the system. The interacting fermions model Hamiltonian projected on a single Landau level in the presence of disorder~\cite{geraedts_17} revealed
signatures of MBL in the presence of magnetic flux but the effects observed were very strongly dependent on the system size. That led the authors to 
postulate the absence of localization in their model in the thermodynamic limit. Apart from these two studies we are not aware of any studies of TRI symmetry breaking effects
in the presence of disorder for interacting systems. The aim of the present work is to fill partially this gap considering experimentally accessible model on a lattice with diagonal disorder.}

 In particular, we examine the spectral and dynamical properties of disordered Hubbard chains in the presence of the synthetic gauge fields. 
At lower disorder strengths, in ergodic regime, the {TRI symmetry} breaking  by gauge field results in spectral statistics 
well described by Gaussian Orthogonal Ensemble (GOE) of random matrices instead of Gaussian Unitary Ensemble (GUE) as expected for broken 
TRI. This is  due to a residual discrete symmetry. Only when the residual reflection symmetry is broken by local field or asymmetric tunneling 
rate of spin-up and {spin-}down fermions, the level statistics is characterized by GUE. {While} the time dynamics of charge and spin correlations for random 
initial states for TRI case reveal the localization of charges and a subdiffusive decay of spin correlations {\cite{prelovsek_16,kozarzewski_18,sroda_19,zakrzewski_18}, the }introduction of a synthetic flux damps 
the spin oscillations and  delocalizes the spins. Furthermore, the entanglement entropy confirms the delocalization of spins in the presence 
of the synthetic gauge fields.

 The paper is organized as follows. In Sec.~\ref{sec:model} we introduce the Hubbard model with synthetic gauge field and disorder. 
In Sec.~\ref{sec:gap_ratio} we analyze the effect of symmetry breaking and synthetic flux on the spectral properties of the 
system. We further study the local charge and spin dynamics of fermions and hard-core bosons in Sec.~\ref{sec:den_corr}. 
In Sec.~\ref{sec:ent_entrop} we examine the bipartite entanglement entropy. The local correlations in the presence of 
spin-dependent disorder are discussed in Sec.\ref{sec:Spin-dependent disorder} followed by remarks on participation ratio of 
eigenvectors { in Sec.\ref{secPR}}.
Finally, we conclude in Sec.~\ref{sec:conc}.

\section{The Model}
\label{sec:model}
\begin{figure}
 \includegraphics[width=1\linewidth]{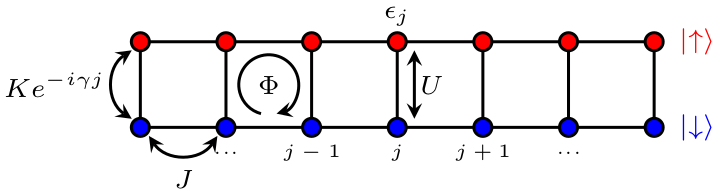}
	\caption{Schematic visualization of the system studied \eqref{model} in which up (down)-spin corresponds to upper (lower) 
	  leg of the ladder. The magnetic flux per plaquette $\Phi = \gamma/2\pi$ \cite{Celi14}.
	  }
\label{picture}
\end{figure}

 We consider interacting spin-1/2 fermions in a quasi-one dimensional lattice. The two spin components can be interpreted as the realization 
of a two-leg ladder geometry, with two legs corresponding to two spin states. Two components may be realized as in the recent 
experiment \cite{schreiber_15} for $^{40}$K atoms with spin up state being 
$|F,m_F\rangle = |\frac{9}{2},-\frac{7}{2}\rangle \equiv |\uparrow\rangle$ and the spin-down state 
$|\frac{9}{2},-\frac{9}{2}\rangle \equiv |\downarrow\rangle$. The two spin components may be coupled by e.g. Raman coupling realizing the 
Hamiltonian \cite{celi_14,suszalski_16} $\hat{H} = \hat{H}_{0} + \hat{H}_{\rm sb}$, wherein
\begin{align}
    \hat{H}_{0} =& \hat{H}_{H}+ \hat{H}_{K} = - \sum_{j, \sigma} \left( J_\sigma~\hat{c}^{\dagger}_{j,\sigma} \hat{c}_{j+1,\sigma} + {\rm H.c.}\right) \nonumber \\
		   & +U \sum_{j} \hat{n}_{j,\uparrow} \hat{n}_{j,\downarrow} 
		   + \sum_{j,\sigma} \epsilon_{j} \hat{n}_{j,\sigma}
		   \nonumber  \\ 
		 & -  K \sum_{j}\left(  e^{-i\gamma j} \hat{c}^{\dagger}_{j,\uparrow} \hat{c}_{j,\downarrow} + {\rm H.c.}\right) .
\label{model}
\end{align}
Here, $j$ and $\sigma = \uparrow, \downarrow$ are the spatial and spin indices, {$J_\sigma$} is the hopping amplitude between neighbouring 
lattice sites on the same leg, $\sigma$, $\hat{c}^{\dagger}_{j,\sigma} 
(\hat{c}_{j,\sigma})$ creates (annihilates) fermions with spin $\sigma$ at site $j$, and the occupation number operator is
$\hat{n}_{j,\sigma} = \hat{c}^{\dagger}_{j,\sigma} \hat{c}_{j,\sigma}$. The local ``charge'' density  is 
$n_j = n_{j,\uparrow} + n_{j,\downarrow}$ while the spin magnetization is $m_j = n_{j,\uparrow} - n_{j,\downarrow}$. The on-site 
interaction strength of two spins is assumed to be repulsive i.e. $U>0$ and $\epsilon_{j}$ is the uniform distributions of a random 
spin-independent on-site potential, $\epsilon_{j}\in[-W/2,W/2]$ with $W$ being the  disorder amplitude. {All these parameters contribute 
to the standard Hubbard part of the Hamiltonian, $H_H$. Note that we allow, in principle, for spin-dependent tunneling rates, $J_\sigma$. Unless explicitly stated we assume, however, that $J_\uparrow=J_\downarrow=J$ recovering the standard disordered Hubbard model.
The second part of the Hamiltonian \eqref{model}, $H_K$ couples two legs of the ladder with   $K$ being  the amplitude and $\gamma$ contributing to  the phase $\gamma j$ of the complex hopping in the synthetic dimension}. $\gamma$ represents the synthetic magnetic flux per plaquette of the lattice. 
 In a laser-induced tunneling scheme (see e.g. \cite{suszalski_16} ) $\gamma = 2k_{R}a$ where
$k_R$ is the recoil wave vector of the laser and $a$ is the lattice spacing. The finite value of $\gamma$ 
leads to  complex hoppings along the rungs of the ladder (corresponding to flipping the spins). {Along the synthetic dimension the system consists of two sites, this might be changed by trapping species in additional Zeeman components, e.g. 
$|F,m_F\rangle = |\frac{9}{2},-\frac{5}{2}\rangle$. We restrict ourselves to the case realized in experiments already \cite{schreiber_15}.} 
The gauge is chosen in such a way that the 
Peierls phase is along the rungs and not on the legs of the ladder - compare Fig.~\ref{picture}.
 The above model Hamiltonian without a synthetic gauge field has been realized in quantum gas experiments in optical 
lattices~\cite{schreiber_15} similarly physics with the synthetic dimension in clean, disorder-free experiments has also been 
studied~\cite{livi_16}.

The hopping amplitude sets the unit of energy scale, $J=1$ and we use periodic boundary conditions throughout this work. We study the system 
at unit filling (the total number of fermions $N = N_{\uparrow}+N_{\downarrow}=L$ is conserved {and equal to a number of sites along the ladder}). Note that such a situation is sometimes 
called a half-filling in condensed matter community. 

{The system concerned has a number of symmetries that have to be taken into account.} In the absence of the gauge field $(K=0)$, the model Hamiltonian preserves parity {(i.e. the exchange between up and down oriented spins)}, 
time-reversal and SU(2) spin symmetry \cite{essler_05}, however, pseudospin SU(2)~\cite{zhang_90} and particle-hole symmetries are broken due 
to the on-site disorder potential. The parity and SU(2) spin symmetry can be removed by adding a local weak magnetic field $(h_b)$ at the edge 
of the chain \cite{mondaini_15}. The symmetry breaking Hamiltonian is
\begin{equation}
    \hat{H}_{\rm sb} = h_b(\hat{n}_{1,\uparrow}-\hat{n}_{1,\downarrow}).
\label{sbreak}
\end{equation}
An alternative approach to obtain the same effect is to allow different tunneling rates for  down-spin fermions $J_\downarrow=J$ and for up-spin fermions $J_{\uparrow} = J + \Delta J$. Such a realization may be simpler in cold atom implementation of the model (it is enough to go away from the so called magic wavelength for lasers forming the optical lattice \cite{schreiber_15}) than creating a local magnetic field. In the absence of $H_K$ that induces coupling between the legs
of the ladder, the number of up-spin, $N_\uparrow$ and that of down-spin,  $N_\downarrow$ fermions are conserved separately. This is no longer true for $K\ne 0$ {when coupling between states with the same parity of $N_\uparrow-N_\downarrow$ is present in the system. } The value of  $N_\uparrow-N_\downarrow$ depends on $N=N_\uparrow+N_\downarrow$;
we have $N_\uparrow-N_\downarrow$ even for $L=6, 8 $ and odd for $L=7$. The conservation of total number of particles implies lack of coupling between the Hilbert space spanned by vectors corresponding to $N_\uparrow-N_\downarrow$ being odd or even. 

{Additionally, the nonzero $\gamma$ - responsible for the synthetic flux - breaks the TRI of the 
{situation studied.} 
Without the 
``symmetry breaking'' terms in the Hamiltonian , e.g. \eqref{sbreak}, this still results in the existence of the so called generalized 
TRI \cite{haake_13} which we study below.}

We use an exact diagonalization for different system sizes and flux quanta obtaining both spectral and time evolution properties of systems 
studied. While using more sophisticated techniques one could extend the study to slightly larger systems (we estimate that the top shift-and-invert technique, \cite{Pietracaprina18} could allow us to reach {$L=10$} with unit filling), we restrict ourselves in 
this exploratory {work} to small $L\leqslant 8$ system amenable to an exact diagonalization. We estimate {the} influence of finite size effects on 
our results by comparison with smaller system sizes. While the question of a possible instability of MBL phase in thermodynamic limit is 
debated~\cite{Suntais19, Abanin19, Sierant19c, Panda19}, we believe that our results are robust on experimentally relevant time scales {of several hundreds of tunneling times and for necessarily finite system sizes in cold atom implementations~\cite{lukin_19,rispoli_19}.}

\section{Spectral statistics}
\label{sec:gap_ratio}
\begin{figure}
 \includegraphics[width=\linewidth]{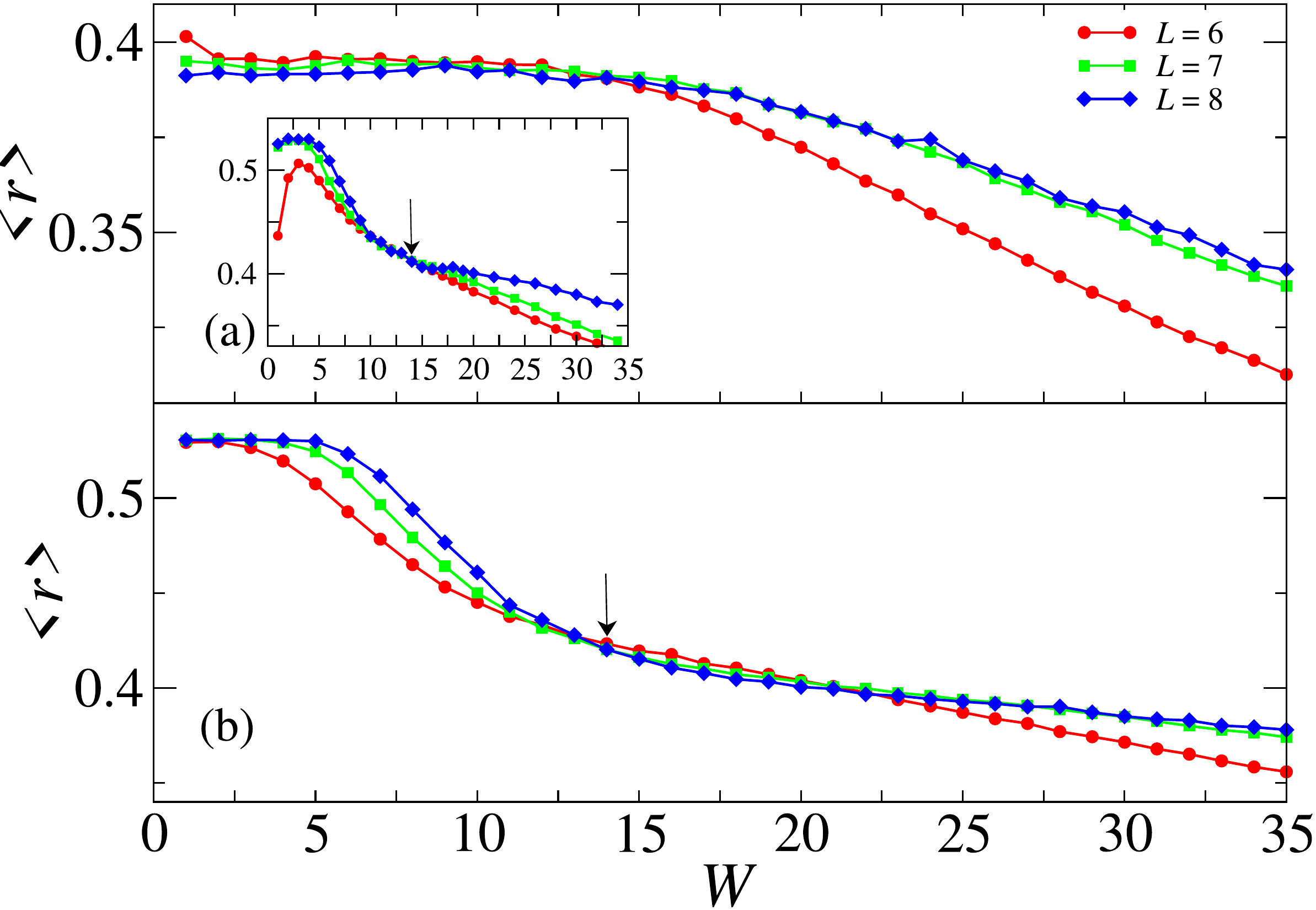}
 \caption{Mean gap ratio $\langle r\rangle$ around the center of the spectra $\epsilon \approx 0.5$ as a function of disorder strength 
	  $W$ for TRI case of $K=1$, $\gamma=0$ of Hamiltonian \eqref{model} in the absence (a) or in the presence (b) of the symmetry 
	  breaking local term \eqref{sbreak} for $h_b = 0.5$. The overlapping distinct spectra in case (a) lead to almost Poissonian level 
	  statistics for any $W$ (main figure). The inset in (a) shows that   diagonalization of a single symmetry block of the Hamiltonian 
	  allow to observe the crossover between GOE-like behavior at low $W$ to Poisson statistics for larger $W$. \
	  The crossing of curves for $L=7$ and $L=8$ (indicated by an arrow) suggests a crossover transition value $W_c\approx 14$.
	   }
\label{sp_rat}
\end{figure}
 
 We consider first the statistical spectral properties of the model. A widely used signature of localization properties of a many-body 
system is the so called gap ratio statistics \cite{oganesyan_07}. From eigenvalue spectrum of the many-body Hamiltonian one computes 
the average gap ratio $\langle r\rangle$, where $r_n = {\rm min}(\delta E_{n}, \delta E_{n+1})/{\rm max}(\delta E_{n}, \delta E_{n+1})$ 
with $E_{n}$ being the energy levels and $\delta E_{n} = E_{n+1} - E_{n}$ is the spacing between two consecutive eigenvalues. It turns 
out that the mean gap ratio may be correlated with the localization properties of eigenstates. For a localized system described by the 
Poisson level statistics, the mean gap ratio is $\langle r_{ \rm Poisson}\rangle = 2 {\rm ln} 2 - 1 \approx 0.3863$; whereas the 
ergodic system is described by $\langle r \rangle$ corresponding to {GOE} $\langle r_{\rm GOE} \rangle = 0.5306$ 
for real Hamiltonian matrix and {GUE} $\langle r_{\rm GUE}\rangle = 0.5996$ in the absence of (the generalized) 
time-reversal symmetry \cite{Atas13}. The mean gap ratio may depends on energy (e.g. for systems with the mobility edge) 
\cite{luitz_15, Nag17, Sheng-Hsuan18, Sierant18} thus it should be determined in the interval of energies where one expects the 
dynamics to be similar. Thus, from now on, we average $r_n$ over the states lying in the center of the spectrum for which 
$\epsilon_n \equiv (E_{n} - E_{\rm min})/(E_{\rm max} - E_{\rm min}) \approx 0.5$ where $E_{\rm min}$ and $E_{\rm max}$ are the 
smallest and largest eigenvalues for a given disorder realization. For $L=6,7,8$ we take {$100,350,500$ eigenvalues} for each disorder realization, respectively.
Finally we average the obtained $r_n$ over disorder 
realizations -- we use over 5000,2000,500 realizations of disorder for $L=6,7,8$ respectively.
\begin{figure}
 \includegraphics[width=\linewidth]{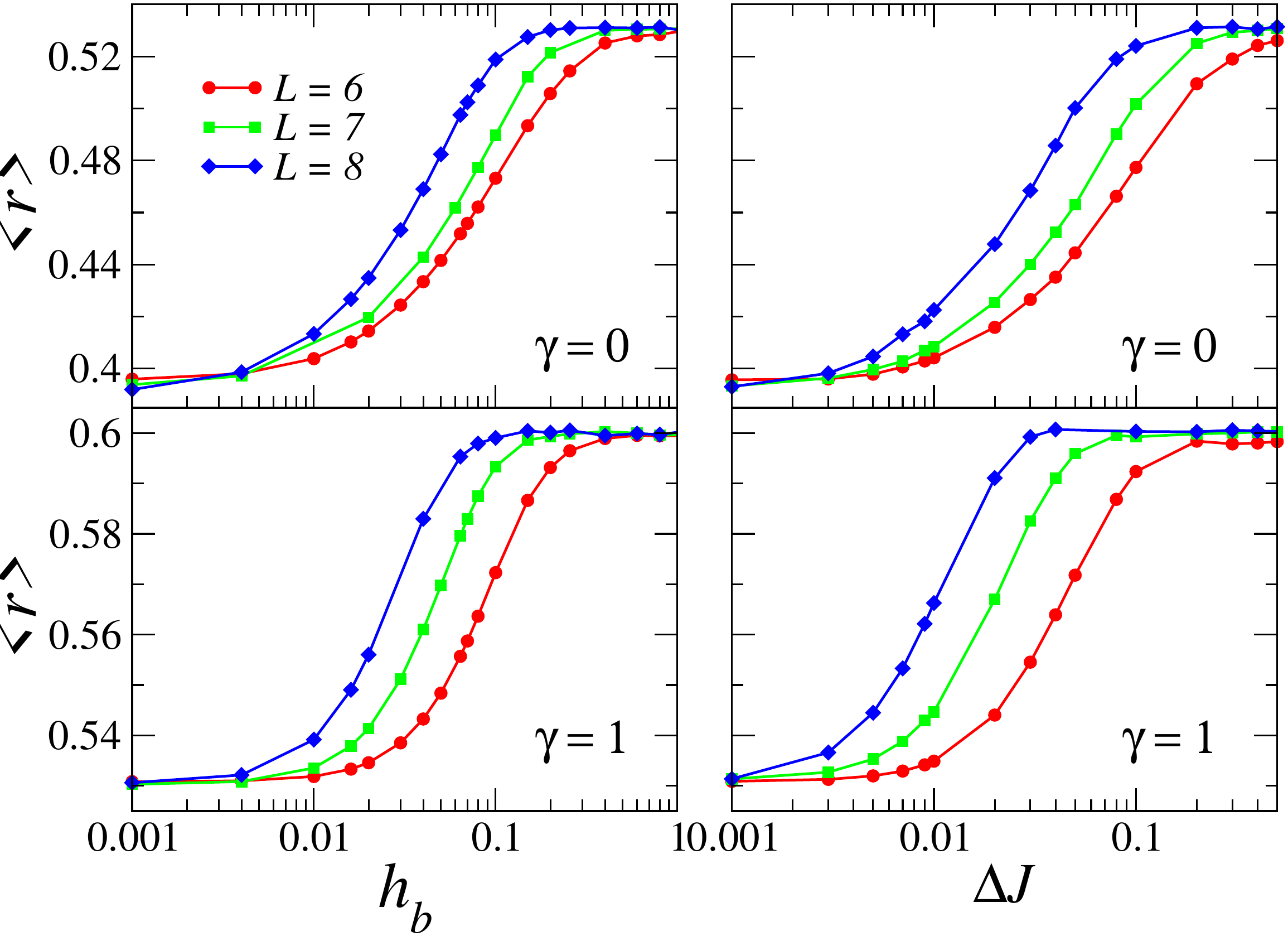}
 \caption{Mean gap ratio for different system sizes in the delocalized regime ($W=3$) as a function of the strength of the 
	  symmetry breaking local magnetic field (left) or the difference between tunneling rates for spin up and down 
	  fermions (right) which also breaks the symmetry. The top row represents TRI case, $\gamma=0$ where presence of 
	  symmetries results in a superposition of independent spectra yielding Poisson-like statistics. For complex fluxes, the 
	  residual reflection symmetry between spin up and down fermions (for unit filling) combined with TRI leads to 
	  generalized time-reversal symmetry (see discussion in the text) leading to an apparent GOE-like behaviour. Only breaking 
	  this symmetry by a local field or difference in tunneling rates, the GUE-like statistics corresponding to broken TRI is observed.}
\label{gr_hb}
\end{figure}

Such a mean gap ratio as a function of disorder strength for Hamiltonian \eqref{model} is shown in Fig.~\ref{sp_rat}. The on-site interaction 
is fixed at $U = 1$. Consider first $\gamma = 0$ case - then TRI in Hamiltonian~\eqref{model} is preserved. Without the symmetry breaking 
term, Hamiltonian~\eqref{sbreak}, the individual spectra from diagonalizations consist of independent subsets of eigenvalues due to conserved 
quantities. Hence, the value of the average gap ratio $\langle r \rangle$ is close to the Poisson value for arbitrary disorder strength $W$ 
vis. [Fig.~\ref{sp_rat}(a)]. To observe the transition between ergodic and MBL phases, we use the explicit forms of generators of SU(2) 
symmetry 
\begin{equation}
   S^z = {\frac{1}{2}}\sum_j ( \hat{n}_{j,\uparrow} - \hat{n}_{j,\downarrow}), 
   	 \quad S^+=(S^-)^{\dag}=\sum_j \hat{c}^{\dag}_{j,\uparrow}\hat{c}_{j,\downarrow}
\end{equation}
\begin{equation}
   S^2 = \frac{1}{2} ( S^+ S^- +  S^- S^+) + (S^z)^2.
\end{equation}
 For $K=1$, the total number of up/down fermions, $N_{\uparrow}/N_{\downarrow}$ is not conserved and $[\hat H_0, S^z] \neq 0$. However, 
the Hamiltonian still commutes with $S^x = (S^+ +S^-)/2$ and $S^2$ operators as can be checked by a direct calculation.  {Finding} 
the Hamiltonian matrix in a basis composed of eigenstates of $S^2$ and $S^x$ and performing exact diagonalization within a single 
block of the matrix, we explicitly observe the crossover between ergodic and MBL regimes as demonstrated in the inset of 
Fig.~\ref{sp_rat}(a).

Alternatively, the crossover can be observed by applying the symmetry breaking local Hamiltonian \eqref{sbreak} \cite{mondaini_15} 
indeed, as Fig.~\ref{sp_rat}(b) demonstrates at weak $W$, $\langle r \rangle$ reaches the value $\langle r_{\rm GOE} \rangle$, while 
at strong disorder it approaches $\langle r_{\rm Poisson} \rangle$. This indicates the existence of two phases at $\gamma = 0$. 

{It would be tempting to extract the critical disorder strength $W_c$ using finite-size scaling approach. We refrain from doing this 
for a few reasons. Firstly system sizes considered are quite small, so extrapolation to infinite system size must be doubtful. Even more importantly,
the finite-size scaling analysis  of MBL transition in spin chains (see e.g. \cite{luitz_15} ) leads typically  to critical exponent $\nu$  
which breaks the Harris bound which is $2$ for one dimensional system \cite{Chandran15a}. The possible explanations may be related to the character of the transition resembling 
a Kosterlitz-Thouless transition \cite{Goremykina19} with important logarithmic corrections. It has been suggested recently that most likely an asymmetric scaling governs the transition,
which may carry over to all observables \cite{Mace19}, see also \cite{lemarie}.}

In view of these problems we tentatively identify the
{disorder strength $W_c$ beyond which the system becomes localized} by the crossing of  $ \langle r\rangle$ curves for largest available system sizes. That leads to $W_c\approx 14$ for $K=1$ $\gamma=0$ i.e. a TRI situation.

\begin{figure}
 \includegraphics[width=\linewidth]{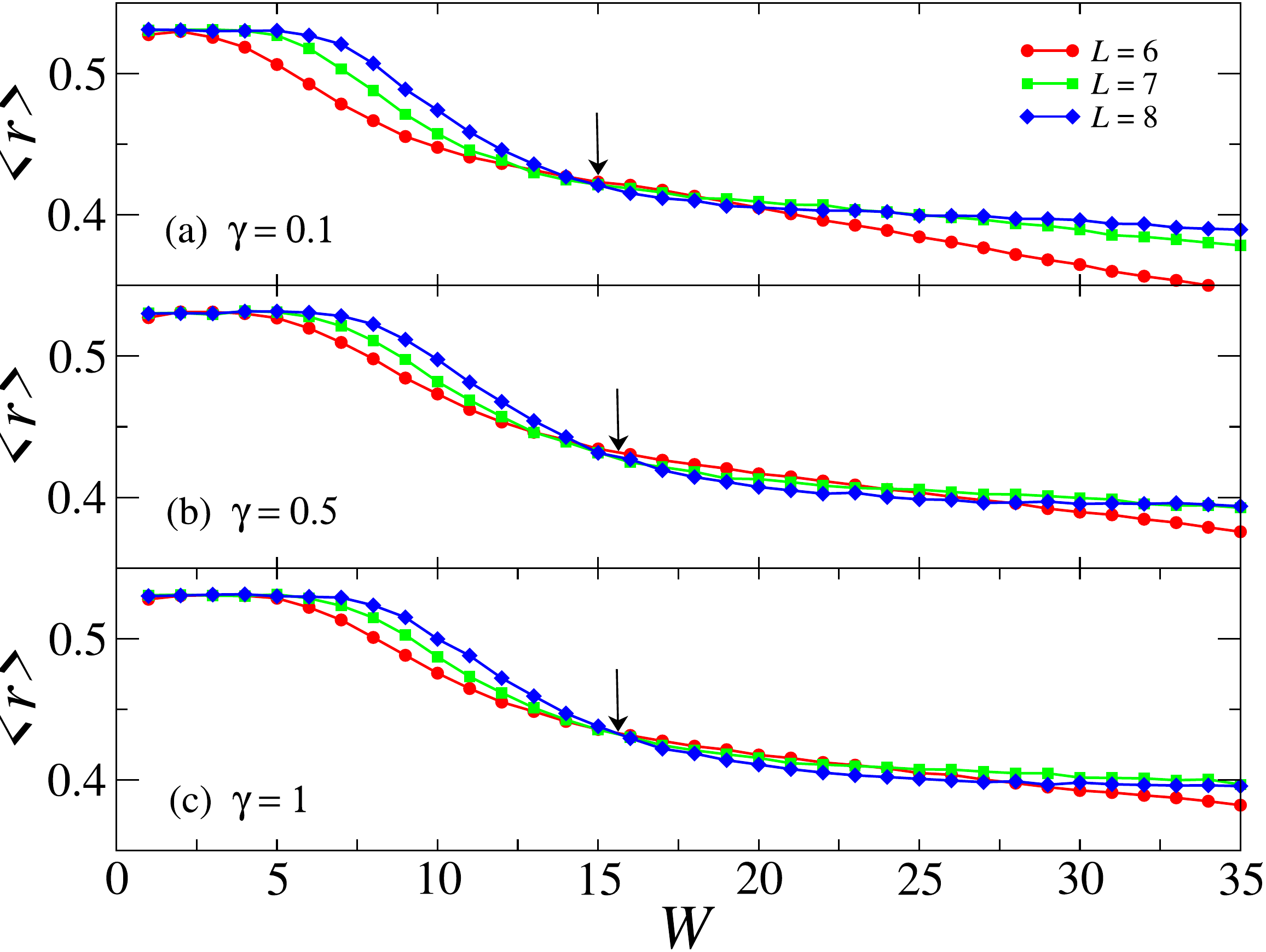}
 \caption{Transition from GOE-like to Poissonian behavior, i.e. from delocalized to localized regime for Hamiltonian \eqref{model} 
	  possessing, for $\gamma>0$, a generalized time reversal symmetry for different values of $\gamma$ as indicated in the 
	  figure. Observe that the crossover point between phases does not depend on $\gamma$.} 
\label{gr_hb0jz}
\end{figure}

The remaining symmetries of Hamiltonian \eqref{model} for TRI case ($\gamma=0$) may be alternatively removed by making the tunneling 
$J$ spin dependent, i.e. $J_{\uparrow} = J_{\downarrow} + \Delta J$. Both cases are visualized in  Fig.~\ref{gr_hb} - top row. 
At $W = 3$ the system is characterized by almost Poissonian mean gap ratio in the presence of symmetries. The finite value of $h_b$ 
couples different symmetry classes and reveals the true extended character of the system with $\langle r \rangle$ corresponding to 
GOE (top left panel). A very similar behavior is observed when spin dependent tunneling is present (top right panel). In both cases the 
larger the system the smaller value of the symmetry breaking parameter is required to fully break symmetry constrains.

 Let us now consider the case of nonvanishing phase $\gamma$ in tunnelings - compare Hamiltonian~\eqref{model}. Non-zero $\gamma$ 
breaks {a standard}  TRI - one might naively expect in that case the GUE-like behavior in {the} delocalized regime. Yet, as shown in the bottom row in 
Fig.~\ref{gr_hb} without $h_b$ (or spin dependent tunneling) the mean gap ratio $\langle r \rangle \approx 0.53$ points towards GOE 
statistics. This is explained by the fact that while standard TRI is broken, there exists a generalized TRI in our system \eqref{model}, 
namely TRI combined with reflection in $x-y$ plane (i.e. change of the spin direction). That generalized TRI leads to GOE statistics 
(for an excellent discussion of symmetries in different universality classes see \cite{haake_13}). 

 With the introduction of $h_b$ or spin dependent tunneling this generalized TRI symmetry is broken and, once the symmetry is fully 
broken, the GUE statistics is fully recovered in the delocalized regime as shown in the bottom row of Fig.~\ref{gr_hb}.

Consider now the crossover from the delocalized to localized regime for nonzero $\gamma$, i.e.,  in the presence of a synthetic 
field. Let us discuss first the case with $h_b=0$ and a generalized TRI - compare Fig.~\ref{gr_hb0jz}. Please note that the observed 
GOE -- Poisson transition in mean gap ratio values seems not to depend on $\gamma$ (once it is sufficiently big to mix different 
{symmetry sectors}) {as we observe examining} the crossing points of
curves in Fig.~\ref{gr_hb0jz} for system sizes $L=7,8$.
\begin{figure}
 \includegraphics[width=1\linewidth]{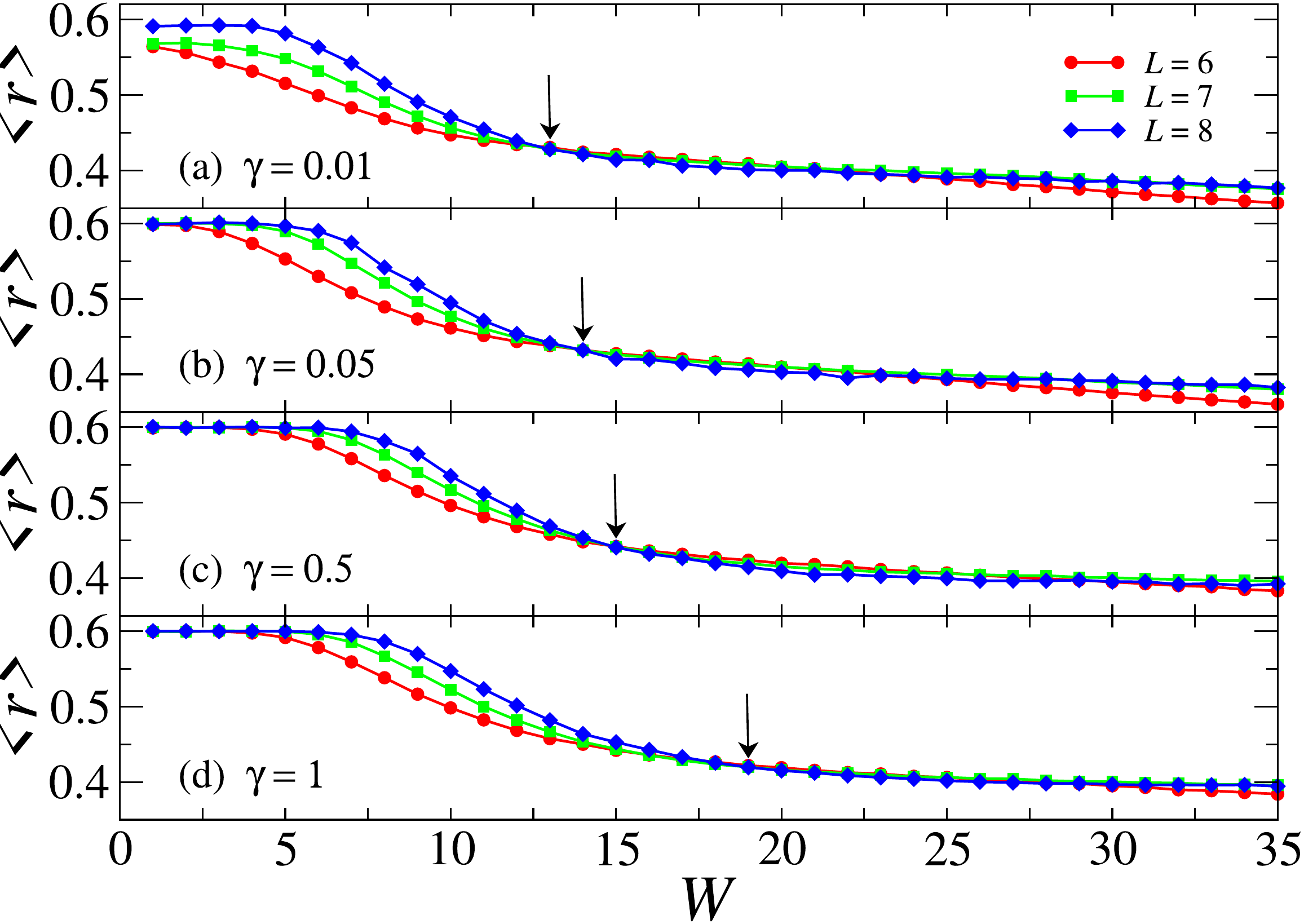}
 \caption{Transition from GUE-like to Poissonian behavior, i.e. from delocalized to localized regime for Hamiltonian \eqref{model} 
	  with additional symmetry breaking term \eqref{sbreak} with $h_b=0.5$. Contrary to $h_b=0$ case presented in 
	  Fig.~\ref{gr_hb0jz} now  the crossover point between phases significantly depends on $\gamma$.  }
\label{gr_hb5jz}
\end{figure}
 
 On the contrary, in the presence of the additional local field [i.e. adding the term \eqref{sbreak} to the Hamiltonian \eqref{model}]
the transition between GUE and Poisson-like behaviour becomes dependent on $\gamma$ as apparent from {data} presented in {Fig.~\ref{gr_hb5jz}}. One might think that this shift of the transition is related to 
the way in which the symmetry is broken, namely by a local magnetic field. Such a local perturbation might differently affect the 
transition  to the localized phase {for different value of $\gamma$}. However, we have checked, by adding random fields on {all of the lattice} sites that this is not the case.

A possible explanation is the following.
{
Consider first the transition between GOE and Poisson level statistics in absence of symmetry breaking field as in Fig.~\ref{gr_hb0jz}. Hamiltonian of the system \eqref{model} for $\gamma>0$ is a complex matrix in the basis of Fock states $| FS_j \rangle$. Constructing a unitary matrix $U_P$ such that 
\begin{eqnarray}
U_P | FS_j \rangle = \frac{ \mathrm{e}^{i \phi_j}}{\sqrt{2}} \left( | FS_j \rangle +  i \mathcal{P} | FS_j \rangle  \right)  \nonumber \\
U_P \left( \mathcal{P} | FS_j \rangle \right)= \frac{ \mathrm{e}^{i \phi_j}}{\sqrt{2}} \left( i | FS_j \rangle +   \mathcal{P} | FS_j \rangle  \right), 
\label{transf}
\end{eqnarray}
{where $\mathcal{P}$ is the parity operator (changing up to down spins and vice versa)}  
we verify numerically that an appropriate choice of phases $\phi_j$ leads to a purely real matrix $H' =  U^{\dag}_P H_0 U_P$. The transformed Hamiltonian matrix $H'$ has the following properties \begin{enumerate}[label=(\roman*)]
 \item diagonal entries of matrix $H_0$ are equal to corresponding diagonal matrix elements of $H'$;
 \item off-diagonal entries of $H_0$ due to hopping along the leg of the ladder correspond to off-diagonal entries of $H'$ of the same number and magnitude;
 \item off-diagonal entries $\mathrm{e}^{i\gamma j}$ of $H_0$ due to hopping in the synthetic dimension become off-diagonal real entries $K\left[\pm \cos(\alpha \gamma)+\sin(\beta \gamma)\right]$ of $H'$ where $\alpha, \beta$ are integer numbers lying in the interval $(-L,L)$.
\end{enumerate}
Fig.~\ref{offdiags} shows distributions $P(m)$ of off-diagonal entries $m$ of the Hamiltonian matrix $H'$. For $1 \ll \gamma  > 0$, the distribution has peaks concentrated at $m=\pm 1, 0$.  
As the value of $\gamma$ increases the distribution $P(m)$ broadens due to 
 the off-diagonal matrix elements associated with tunnelling in the synthetic dimension (iii).
 For values of $\gamma$ larger than the certain system size dependent $\gamma_0$, the distribution $P(m)$ changes only mildly with $\gamma$. Since the diagonal entries of the matrix remain intact, it seems plausible that the independence of the
crossover point of the value of $\gamma$ observed in Fig.~\ref{gr_hb0jz} is caused by the fact that the distribution of off-diagonal entries does not change significantly once $\gamma \gtrsim 0.1$ for system size $L=8$.
\begin{figure}
 \includegraphics[width=1\linewidth]{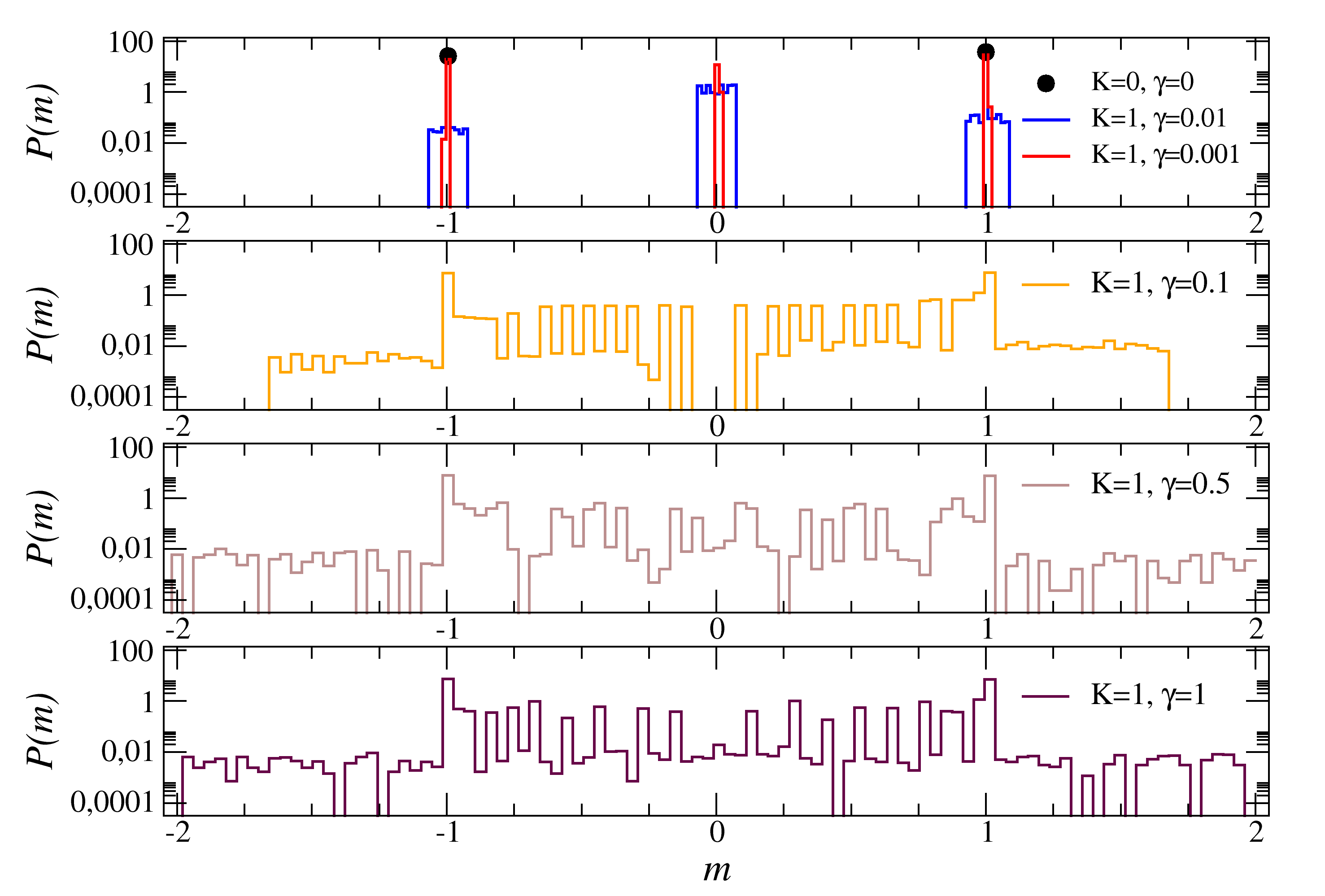}
 \caption{ {The distribution of off-diagonal entries of matrix $H'$ for various values of parameters $K$ and $\gamma$ for the system size $L=8$. Once $\gamma$ is larger than $\approx 0.1$ the distribution is only  weakly dependent on the value of  $\gamma$. } }
\label{offdiags}
\end{figure}
}

{The properties of the crossover in the presence of the additional symmetry breaking field, e.g. $h_b=0.5$ as in Fig.~\ref{gr_hb5jz}, are starkly different -- the position of crossover shifts with $\gamma$ especially for $\gamma > \gamma_0$. Viewing $H_{sb}$ \eqref{sbreak} as a perturbation to eigenstates of $H_0$, we see that even though the overall properties of the unperturbed system change with disorder strength $W$ independently of $\gamma$ (provided that $\gamma>\gamma_0$), the eigenstates depend on the value of $\gamma$. This conclusion is further corroborated in the study of eigenstates properties in Section~\ref{secPR}.}
\section{Dynamics and density correlations}
\label{sec:den_corr}

To study the dynamical properties, we choose random Fock state $|\psi(0)\rangle$ as an initial state for the temporal evolution 
and examine the dynamics of the local correlations and the entanglement entropy for $L=8$. The local charge and spin correlations are 
${C(t)} = A \sum_{j}\langle\rho_j(t)\rho_j(0)\rangle$ and ${S(t)} = B \sum_{j}\langle m_j(t)m_j(0)\rangle$, 
where $\rho_j = n_j - \bar{n}$ with the average density $\bar{n} = \sum_j n_j/L$, and $A$ and $B$ are normalization constant such 
that ${C(0)} ={S(0)} = 1$. Recall that $n_j$ ($m_j$) are the sum (difference) {of the} site occupations of up and down 
polarized fermions, respectively. We assume a unit filling, $\bar{n} = 1$. 

Let us recall what is known about the disordered Hubbard case, $K=0$. The memory of the initial state is lost for sufficiently small disorder. {The system is delocalized} and this leads to the decay of charge and spin correlations. An interesting situation occurs  for stronger disorder where (in the 
absence of symmetry breaking local magnetic field or other effects destroying the system symmetries) the charge sector appears 
localized but spin sector does not show localization~\cite{prelovsek_16,zakrzewski_18}. The absence of a fully localized phase is due to 
the SU(2) symmetry of the model. The choice of a different random potential for up and down polarized fermions, a random magnetic 
field or a weak spin asymmetry which breaks the SU(2) spin symmetry recovers the full MBL 
phase~\cite{lemut_17,sroda_19,Leipner-Johns19}. For Hubbard model the subdiffusive transport of spins can be traced back to a singular random distribution 
of effective spin exchange interactions~\cite{kozarzewski_18} and at long times the transport is strongly 
suppressed~\cite{zakrzewski_18}. The particle transport rate is exponentially small in $J/W$, and the rate depends strongly on the 
initial state of the system. The states occupied with only doublons or holons exhibit full MBL phase as for such states the 
delocalization rate is zero~\cite{protopopov_19}. This is the current understanding of dynamical properties of disordered Hubbard 
model.

\begin{figure}
 \includegraphics[width=\linewidth]{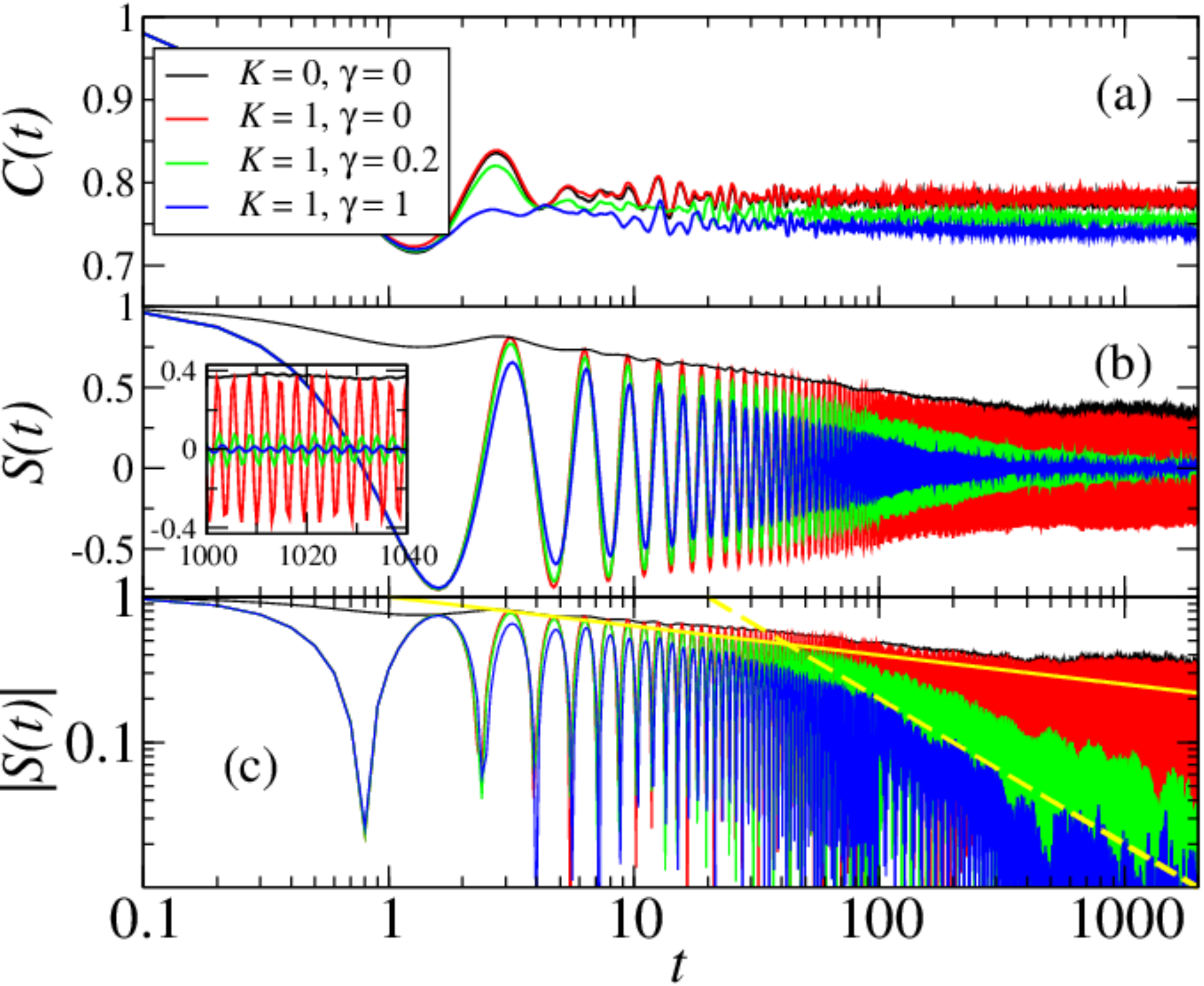}
 \caption{Time evolution of the local charge and spin correlations of random Fock states for $U=1$, $W=32$ and $h_b=0$. (a) The charge 
	  remains localized in the presence of the synthetic flux. (b) In the absence of the flux, when $K=0$, the spin delocalizes with 
	  subdiffusive decay of correlations (see text for discussion). Once $K$ is finite, the spins oscillate (red curve) and the $S(t)$ of $K=0$ (black curve) 
	  provides an envelope to these oscillations. This is apparent from the inset plot where the oscillations in the evolution at large times are {shown}. At $\gamma>0$, the oscillations are damped and $S(t)$ decays to zero. The gauge field delocalizes the spins.  Panel (c) - the modulus of $S(t)$ in the log-log scale. Yellow line proportional to $1/t^{0.2}$ shows the subdiffusive character of decay for $K=0$. The decay becomes faster and for $K=\gamma=1$ becomes ballistic-like with $1/t$ behaviour as exemplified by a dashed yellow line.
	  The average is performed for $300$ disorder realizations.}
\label{corr_ferm}
\end{figure}

 Considering the coupling of spin up and down particles allows for further studies of ergodicity breaking in the model. We concentrate mostly on the strongly disordered regime  where either charge or spin correlations decay very slowly on the time scale considered - we study the dynamics up to $t=2000/J$ which seems to be at the edge of current experimental possibilities with cold atomic setups.  The time evolution of density and spin correlations are shown in 
Fig.~\ref{corr_ferm}. We examine the behaviour of  correlations at several values of the synthetic flux to observe its influence on 
the dynamics. For $\gamma=0$, $K=0$ (i.e. no coupling between the up and down polarized fermions) the charge correlation $C(t)$ 
saturates at  a finite value for an exemplary  $W=32$ value. The spin correlation $S(t)$ shows a slow 
decay claimed in earlier works to be due to subdiffusive spin transport~\cite{prelovsek_16,zakrzewski_18,kozarzewski_18,sroda_19}. Fig.~\ref{corr_ferm}(c) shows the decay of spin correlations in the log scale, observe that the decay up to time $t\approx300$ follows a power law, $|S(t)|\propto 1/t^{0.2}$ then the slope of the curve diminishes. It is worth noting that a similar behavior was observed for a much larger system in \cite{zakrzewski_18}.  

 Also at zero flux ($\gamma=0$), the coupling of spin degrees of freedom 
through finite hopping along the synthetic dimension ($K=1$) does not alter the behaviour of $C(t)$ and the charge remains localized on the timescale considered. 
In the presence of a finite synthetic flux $\gamma=0.2$ and $1$, after a transient time of the order of $1/J$, the charge 
again seems to saturate. Yet a careful analysis of $C(t)$ reveals a residual very slow algebraic decay, $C(t)\propto t^{-\alpha}$ with 
$\alpha=0.004(2)$ for $K=\gamma=1$ as compared to order of magnitude less $\alpha$ for $\gamma=0$. Since this power is quite small we have performed detailed tests
using bootstrap technique to analyse the error (1150 disorder realizations are used). $\alpha=0.004(2)$ is obtained when fitting the decay over  $t\in [100,200]$. Fitting the slope we observe that $\alpha$ decreases with time, e.g. $\alpha=0.0021(3)$ in the interval $t\in [200,1000]$ suggesting that the decay is slower than the power law.
Also,  $\alpha=0.004$ is significantly smaller than the exponent governing residual decay of imbalance at MBL transition 
in large systems of spinless fermions \cite{Doggen18,Chanda19} thus  
we believe that a small value of $\alpha$ indicates that the charge sector of {spinful} fermions remains localized in the presence of  the synthetic gauge field for a 
sufficiently strong disorder. Let us also note that the power of the small residual decay is also  dependent on the system size increasing with it, e.g. we get $\alpha=0.0032(7)$ for $t\in[100,200]$ for $L=7$.

Consider now the evolution of the spin correlation $S(t)$ which is shown in Fig.~\ref{corr_ferm}(b). In the absence of the flux, 
{$S(t)$ decays slowly in agreement with the subdiffusive manner suggested in such a case \cite{zakrzewski_18,kozarzewski_18,sroda_19}.
As in the study of much larger system sizes \cite{zakrzewski_18}  for times up to around $t=300-350$ the decay is faster with a power law characted $t^{-0.175(1)}$
slowing down at large times to  $t^{-0.037(2)}$. As discussed in \cite{zakrzewski_18} the functional dependence of the decay in $t\in(300,2000)$ interval is hard
to be determined numerically (equally well one might assume a logarithmic dependence). Powers of decay slower than $1/2$ suggest a subdiffusive behavior 
\cite{Luitz17} as proposed \cite{kozarzewski_18,sroda_19}. }

{The presence of coupling between the legs of the ladder $K>0$ strongly affects the spin dynamics - $S(t)$ becomes oscillating. }
The frequency of oscillations is 
simply $K/2$ as may be verified inspecting the inset of Fig.~\ref{corr_ferm}(b). 
Interestingly, the oscillations have the envelope given by the 
$S(t)$ curve for $K=0$. This quite surprizing at first observation may be easily explained.
Denote by $H_K$ the term of $H_0$ proportional to $K$ and by $H_H=H_0-H_K$.  We verify that 
$[H_H, H_K]=0$ as long as $\gamma=0$. Thus, the evolution operator factorizes: $\exp(-i H_0 t) = \exp(-i H_H t) \exp(-i H_K t)$ and 
average number of fermions with spin $\sigma$ at site $j$ is given by 
\begin{equation}
  n_{j\sigma}(t) = \langle\psi(0) | e^{iH_H t}e^{iH_K t} \hat{n}_{j\sigma}e^{-iH_K t}e^{-iH_H t} | \psi(0)\rangle,
\end{equation}
so that the time evolution of spin correlation function is a superposition of  dynamics determined by the $H_H$ Hamiltonian and 
oscillations driven by the $H_K$ term. The long-time average of the oscillations is zero, which is in agreement with previous study 
on the localization of coupled chains with correlated disorder~\cite{zhao_17}. 

In the presence of a finite flux, $[H_K, H_H]\neq0$, 
the oscillations are damped and the damping rate grows with $\gamma$. Thus, the spin sector of the system gets delocalized by 
the introduction of gauge field.  The decay of spin correlation function changes from the subdiffusive at $\gamma=0$ continuously 
reaching $1/t$ behavior at $\gamma=1$ characteristic for a ballistic-like motion \cite{Luitz17} - compare Fig.~\ref{corr_ferm}(c). 
\begin{figure}
  \includegraphics[width=\linewidth]{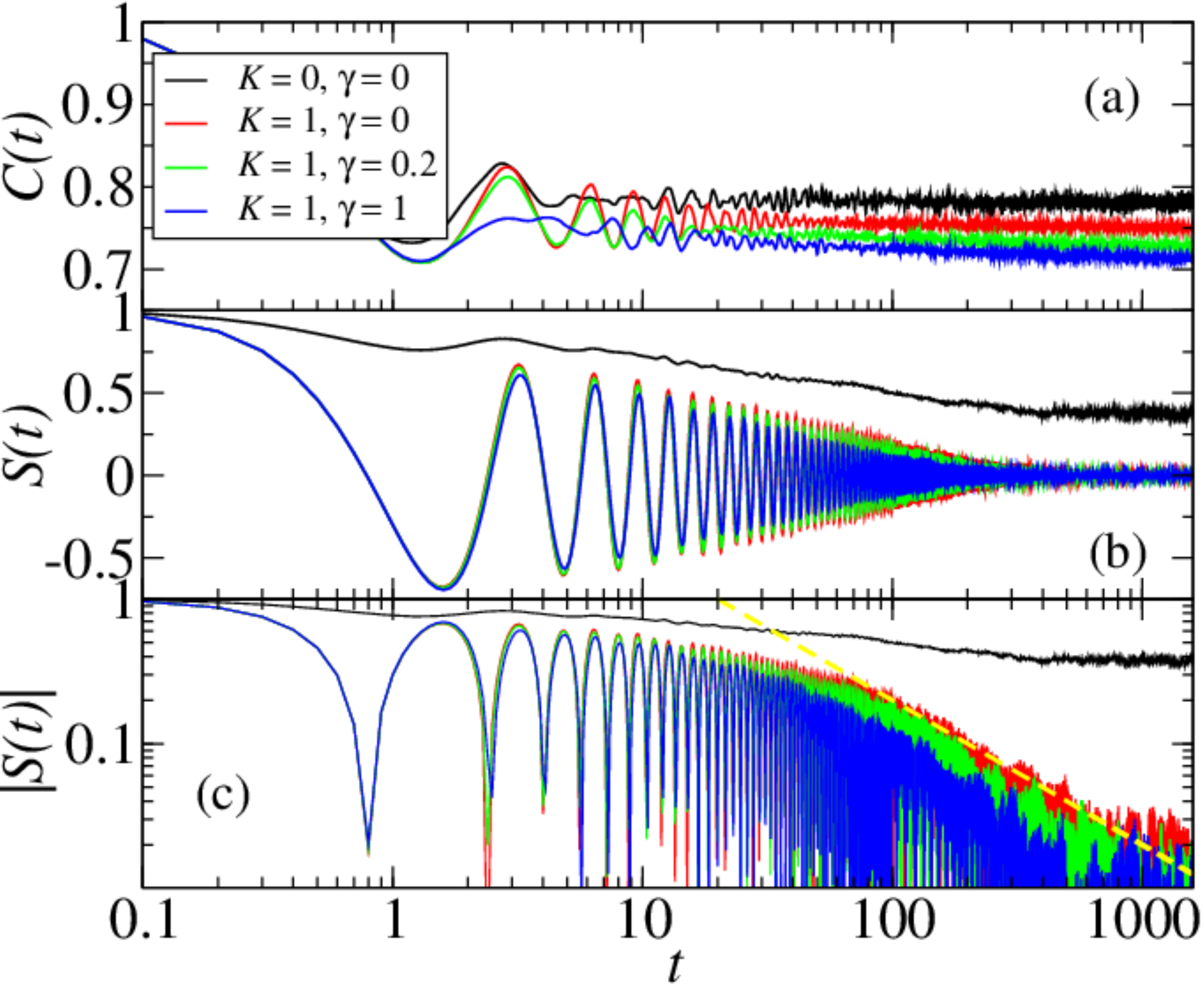}
  \caption{The local correlations of the hard-core bosons in the localized regime $W = 32$. The initial states are the random Fock states 
	   and $U=1$. (a) The charge remains localized in the presence of gauge field. (b) Once the tunneling coupling $K$ between up and down spins is turned on the spin correlations decay to zero showing transient
	   oscillations with $K/2$ frequency. The decay of the envelope seems to be ballistic-like at longer times -- a dashed bright (yellow on-line) line in panel (c) indicates $1/t$ decay.
	   }
\label{corr_hcb}
\end{figure}

 To examine the effect of quantum statistics, we further analyze the time dependence of local correlations when instead of fermions we 
consider hard-core bosons in the Hamiltonian \eqref{model}. A cold atom implementation might use $^{39}$K instead of $^{40}$K also with Zeeman sublevels. Using well controlled Feshbach resonances \cite{Errico07} one may increase interactions between bosons to a degree when double occupations in the lattice become energetically prohibited realizing hard-core boson limit. The correlations of hard-core bosons for different values of $K$ and 
$\gamma$ are shown in Fig.~\ref{corr_hcb}. For $K=0$ i.e. no phase-sensitive coupling between the legs of the ladder both charge and spin correlations
behave similar to the fermionic case. The situation changes for $K>0$. The behaviour of charge correlation, $C(t)$, is similar to the fermionic case and its finite 
saturation value indicates the localization of the charge sector in the presence of the synthetic flux. The saturation value of $C(t)$ 
decreases with $0\leqslant\gamma\leqslant1$ at $K=1$. The spin sector behaves differently.  The coupling of  two hyperfine states (along the rungs of the ladder), even for $\gamma=0$, leads to  oscillations of spin correlations, which in contrast 
to fermionic $S(t)$, are damped to zero. { This is due to the fact that the commutator $[H_K, H_H]\neq0$ for hard-core bosons
leading to the damping of  $S(t)$.
}
A finite flux $\gamma$ only weakly affects the damping and spin sector becomes fully 
delocalized. The decay of spin correlations seems almost exponential at first. Then, at later times, as indicated by yellow dashed line in fig.~\ref{corr_hcb}(c), the decay of spin correlations is ballistic-like, i.e. governed by $1/t$ behavior independently of the flux $\gamma$.
 The stark contrast of $S(t)$ between hard-core bosons and fermions for $K=1$ and $\gamma=0$ is a very nice demonstration 
of effects induced by quantum statistics and different commutation properties of fermions and hard-core bosons 
(for ground state properties this observation goes back to \cite{Crepin11}).

\begin{figure}
  \includegraphics[width=\linewidth]{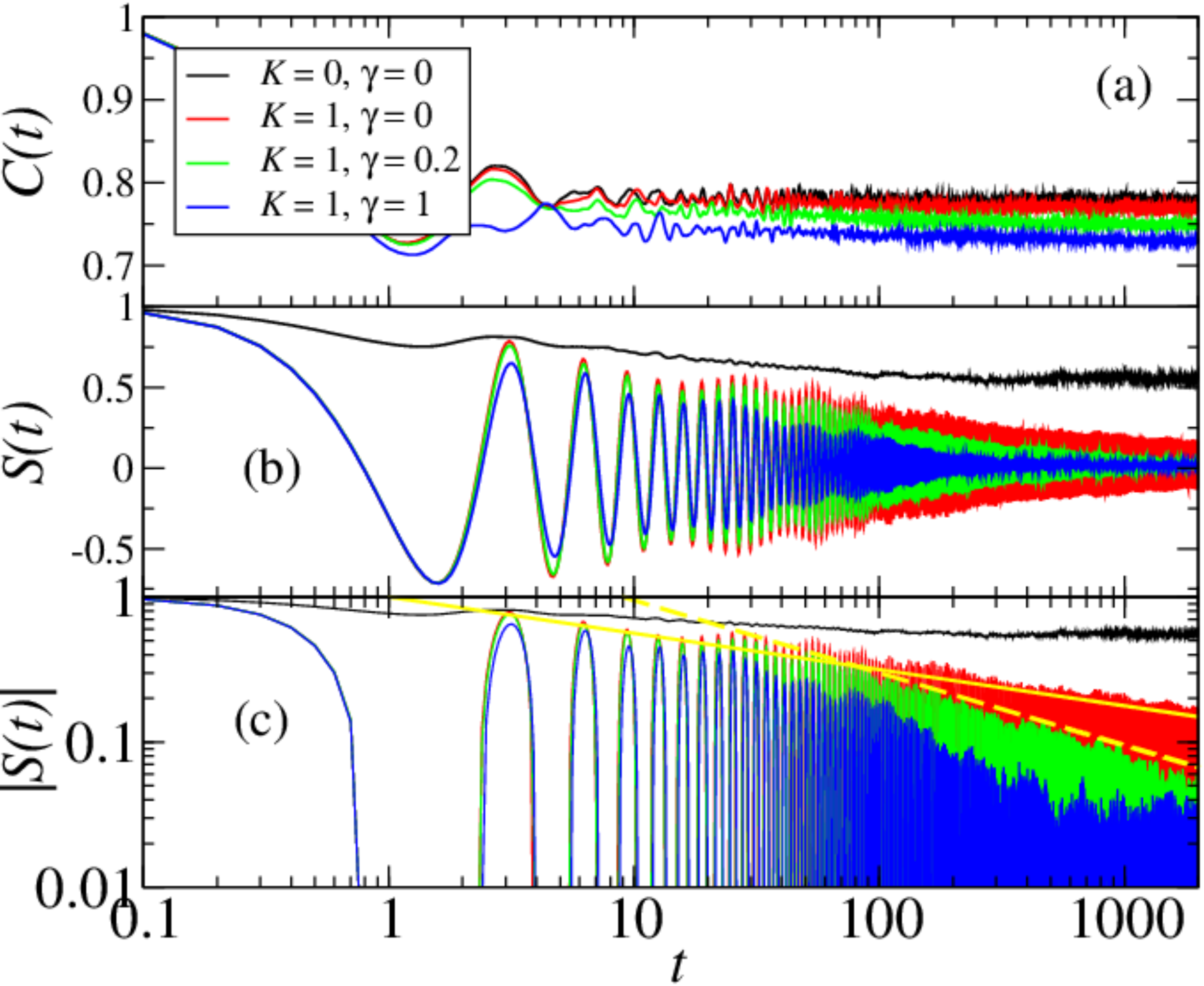}
  \caption{Same as Fig.~\ref{corr_ferm} but in the additional presence of the local magnetic field term, [Eq.\eqref{sbreak}] breaking the 
	   generalized TRI symmetry.  
	   The strength of local field $h_b = 0.5$. The presence of coupling $K$ induces faster decay of $S(t)$ as compared to the case without symmetry breaking local field (Fig.~\ref{corr_ferm} - still the decay is subdiffusive as indicated by a solid bright (yellow) line with slope $1/t^{0.25}$. In the presence of the flux the decay is faster, approximately diffusive for $\gamma=0.2$ - see the yellow dashed line proportional to $1/\sqrt(t)$, for larger $\gamma$ becoming even faster. }
\label{corrp5}
\end{figure}

Finally, we return to the fermionic system and calculate time evolution of charge and spin correlation functions as shown in 
Fig.~\ref{corrp5} for non-zero value of symmetry breaking field $h_b$. In this case, even at $\gamma=0$ the commutator 
$[H_H + H_{\rm sb}, H_K] \neq 0$ and the oscillations of $S(t)$ are damped. The effects of broken generalized TRI symmetry, while altering 
significantly spectral properties of the system, have relatively small effect on time dynamics of the charge correlation function $C(t)$ for 
$\gamma>0$. On the other hand, for spin correlation function, $S(t)$ we again observe a transition from a subdiffusive (at nonzero $K$ and zero flux) to superdiffusive (for sufficiently large flux $\gamma$) decay of the spin correlation function.

\section{Entanglement entropy growth}
\label{sec:ent_entrop}

 To corroborate our study of spin delocalization in the presence of the synthetic gauge field we investigate the entanglement entropy 
growth in the system. The two-leg ladder system can be partitioned in several ways with two prime options: the first {one} is to cut 
the ladder perpendicular to the direction of the spatial dimension in two equal (left-right) parts and the second {one} is to cut 
the ladder perpendicular to the direction of synthetic dimension decoupling the hyperfine states. This allows us to study both the 
entanglement between two left and right ladder subsystems but also between an ensemble of up and down polarized fermions.

Regardless of the splitting the von Neumann entanglement entropy is defined in a standard way as
\begin{equation}
  S_E(t) = - \Tr \rho_A(t) \ln \rho_A(t),
\end{equation}
where the two subsystems are denoted as $A$ and $B$ with $\rho_A(t) = \Tr_B
{\ket{\psi(t)}\bra{\psi(t)}}$ being the reduced density 
matrix of subsystem $A$. Here $\Tr_B$ is the trace over degrees of freedom of subsystem $B$. 
\begin{figure}
  \includegraphics[width=\linewidth]{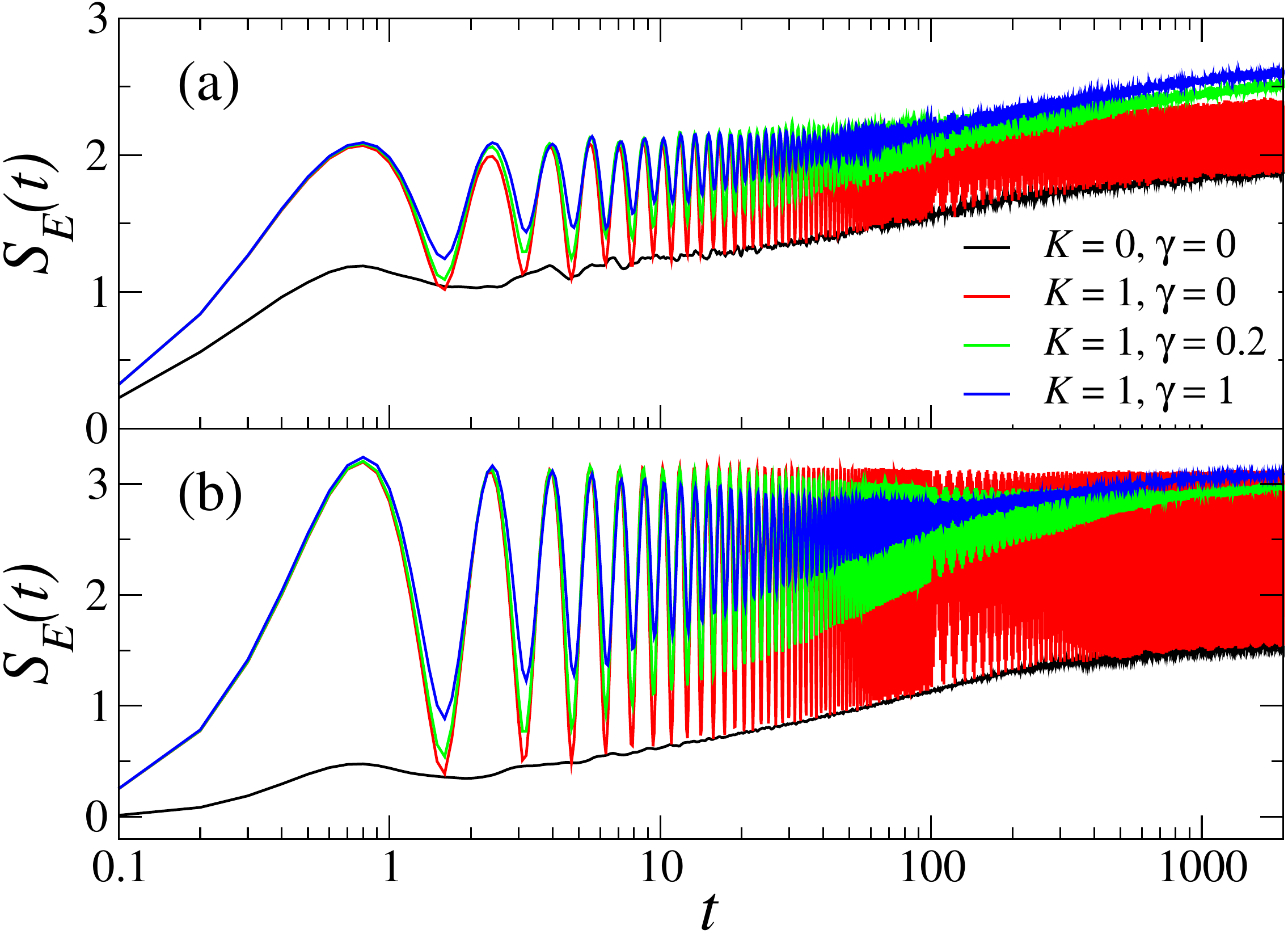}
  \caption{(a) Left-right bipartite entanglement entropy in the localized regime. With no coupling of spin up and spin down polarized 
	   fermions a typical logarithmic growth of $S_E$ is observed after initial transients. Coupling $K$ induces oscillations 
	   of the entanglement entropy. Complex flux $\gamma$ leads to damping of these oscillations with also faster entropy growth.
           (b) Bipartite entanglement entropy dynamics when rungs of the ladder are cut, i.e. a partition divided up and down polarized 
	   fermions.}
\label{lr_ud_ent}
\end{figure}

 In the MBL regime the entropy of entanglement of initially separable state should grow logarithmically in 
time \cite{bardarson_12,Serbyn13a}  {after an initial transient eventually saturating for a finite system size \cite{Serbyn13a}.}
 Such a behavior is indeed observed in our model for decoupled spin chains ($K=\gamma=0$) both for the standard left-right 
splitting [Fig.~\ref{lr_ud_ent}(a)] and for the up-down splitting [Fig.~\ref{lr_ud_ent}(b)].

The coupling between up and down polarized fermions introduces strong oscillations in the entanglement entropy, with the same period 
as revealed by the spin correlation function - compare Fig.~\ref{lr_ud_ent}(a). Similarly, the entropy curve for $K=0$ forms an 
envelope (here low lying envelope) to the oscillations. Again this fact is related to vanishing commutators between 
different kinetic energy terms in the Hamiltonian \eqref{model}, as described in the previous Section. The presence of complex flux makes 
these commutators non-zero and introduces the damping of the oscillations. The resulting entropy growth is significantly faster than in the 
$K=\gamma=0$ case reflecting the delocalization of spin sector.

Similar, even more spectacular oscillations of the entanglement entropy are observed when the partition divides up and down polarized 
fermions (i.e. the rungs in the ladder are cut). The coupling between up and down fermions ($K=1$) makes the entropy oscillate 
maximally, with lower envelope given by $K=0$ curve as before while the upper envelope is system size and disorder strength dependent. 
Again the non-zero flux introduces damping of these oscillations - on the time scale discussed the entropy growth practically saturates 
at upper allowed value [Fig.~\ref{lr_ud_ent}(b)].

\section{Spin-dependent disorder}
\label{sec:Spin-dependent disorder}
\begin{figure}
  \includegraphics[width=\linewidth]{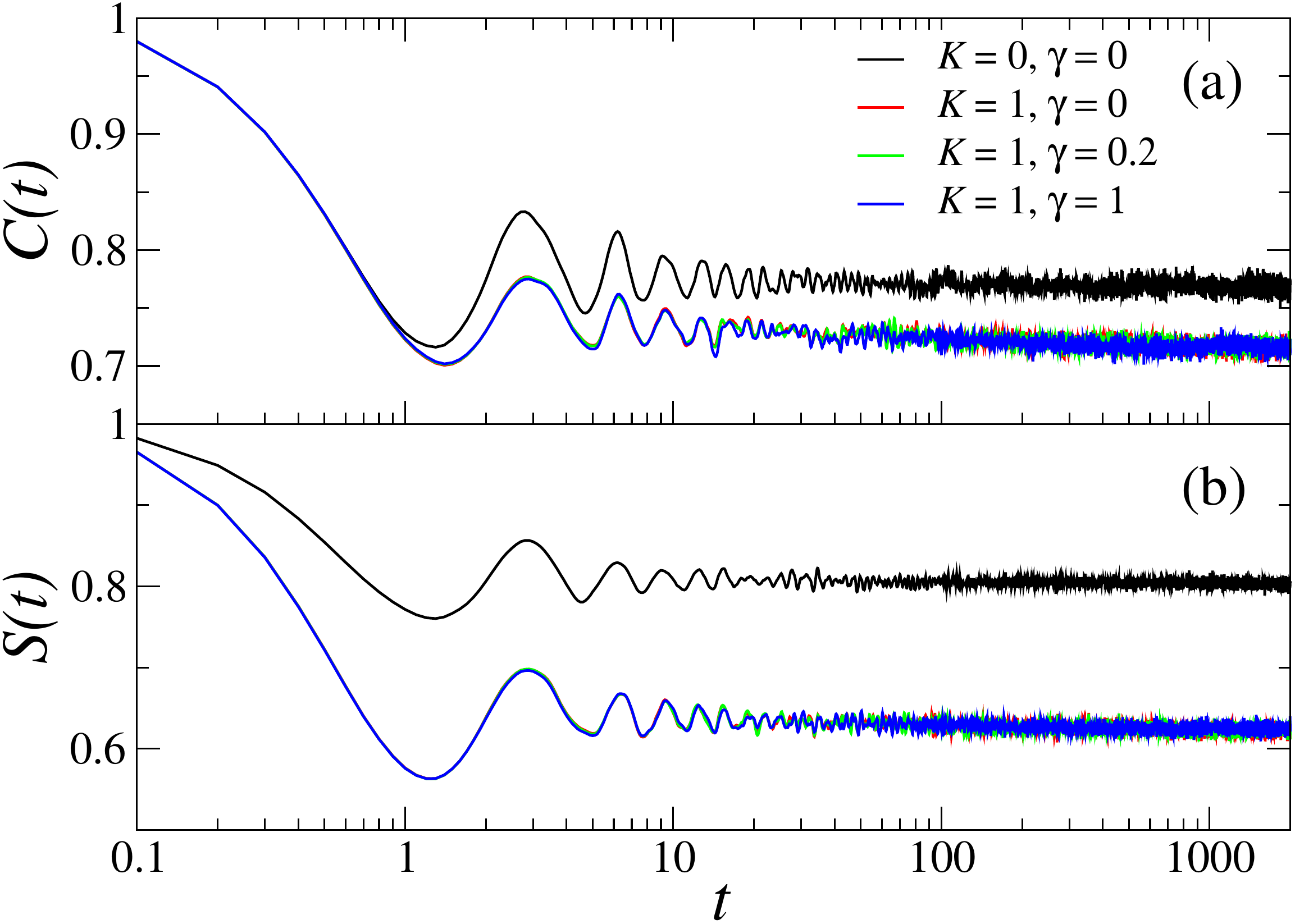}
  \caption{(a) Charge $C(t)$ and (b) spin $S(t)$ correlation functions for uncorrelated disorder $\epsilon_{j\sigma} \in [-W/2, W/2]$, 
	   the chemical potential is site and spin dependent. The system size is $L=8$, disorder strength $W=32$, various values of 
	   coupling $K$ and flux $\gamma$ are considered. Both $C(t)$ and $S(t)$ saturate signalling localization of both charge and 
	   spin sectors.}
\label{corr_unc}
\end{figure}

 Up till now we {have} considered the disorder identical for both spin components. Such a situation (albeit for quasiperiodic disorder) is 
realized in experiments \cite{schreiber_15}. However, it is possible (and feasible) experimentally 
 to consider a situation when disorder is different 
(and uncorrelated) for both spin components {with}   the last term in $\hat{H}_{H}$ of Hamiltonian \eqref{model} {being} changed to 
$\sum_{j,\sigma} \epsilon_{j,\sigma} \hat{n}_{j,\sigma}$ where now $\epsilon_{j,\sigma} $ are independent random variables drawn from 
uniform distribution in the interval $[-W/2,W/2]$. The resulting charge and spin correlation functions are 
shown in Fig.~\ref{corr_unc}. We observe a clear localization of both charge and spin sectors as $C(t)$ and $S(t)$ saturate to constant values 
after initial oscillations. The fact that addition of a (small amount of) disorder in spin sector, in presence of strong charge 
disorder is sufficient to fully localize the system was also observed in \cite{Leipner-Johns19}. The saturation values are lower for 
$K=1$ than for $K=0$ and, interestingly, are independent of the flux $\gamma$. 
\begin{figure}
  \includegraphics[width=\linewidth]{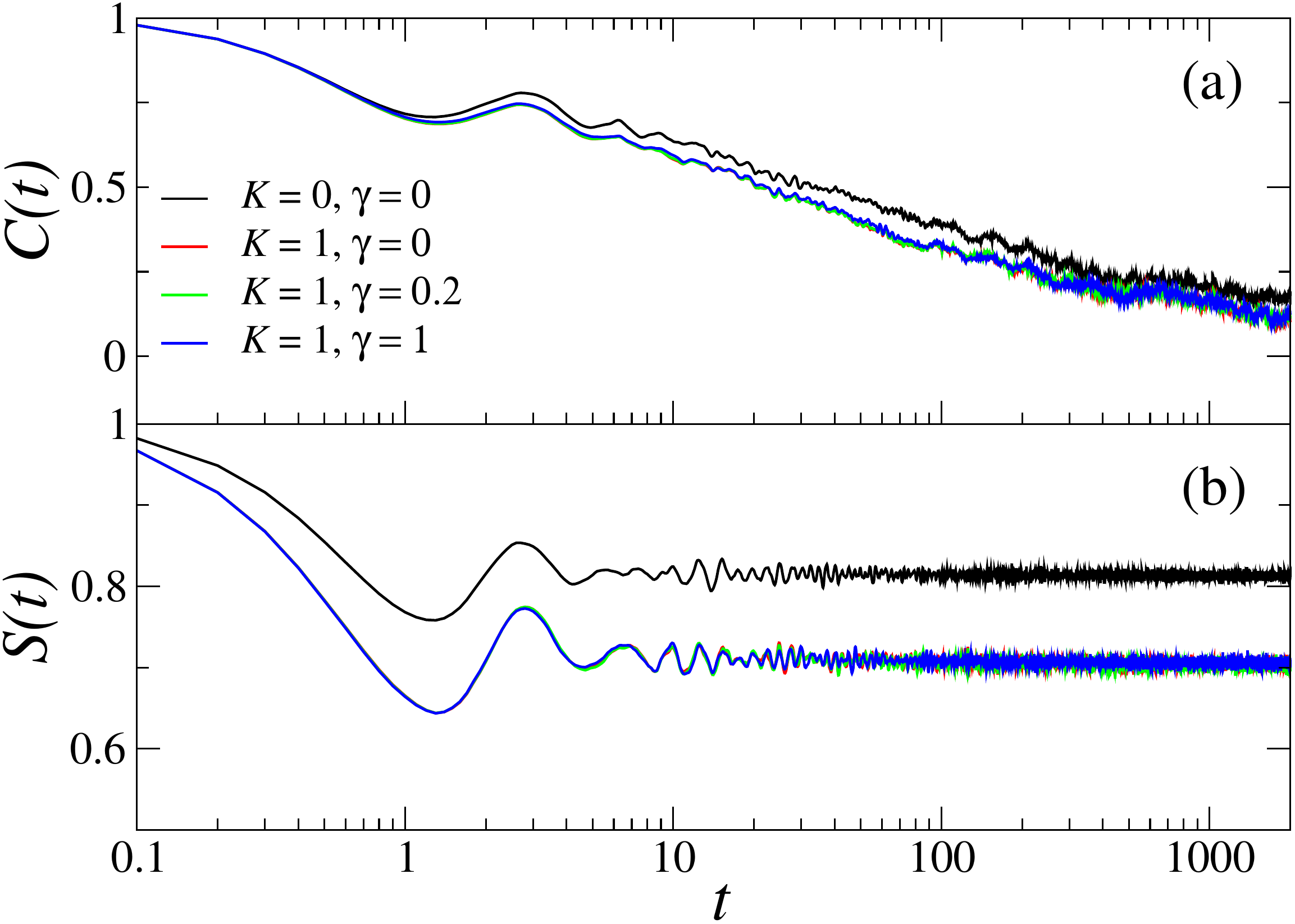}
  \caption{(a) Charge $C(t)$ and (b) spin $S(t)$ correlation functions for spin disorder $\epsilon_{j\sigma} =-\epsilon_{j\bar \sigma}$ 
	   (where $\bar \sigma$ denotes spin opposite to $\sigma$). The system size is $L=8$, disorder strength $W=32$, various values 
	   of coupling $K$ and flux $\gamma$ are considered. The saturation of $S(t)$ signals localization of the spin sector whereas 
	   the charge sector remains delocalized.}
\label{corr_spin}
\end{figure}

 To complete the picture we consider the case of  ``spin disorder'' namely, we replace the last term in {$\hat{H}_{H}$ of}
 Hamiltonian~\eqref{model} by 
$\sum_{j} \epsilon_{j} (\hat{n}_{j,\uparrow}- \hat{n}_{j,\downarrow})$ so the disorder affects the spin degree of freedom. 
{ In this case, for $K=0$, the system possesses pseudo-spin SU(2) symmetry that can be utilized in studies of its localization 
properties \cite{Yu18}. Charge and spin correlation functions for various values of $K$ and $\gamma$ are shown in 
Fig.~\ref{corr_spin}. We observe that the spin sector is localized as $S(t)$ quickly saturate whereas the charge correlations decay 
suggesting delocalization of charge degrees of freedom.}

Note that when disorder becomes spin-dependent the time evolution of both charge and spin correlation function is no longer 
dependent on the phase of the coupling $\gamma$ between spin components. While for individual realizations of disorder the evolution may differ
substantially, the differences average out when taking the disorder average. Once the spin degree of freedom becomes localized the 
information transfer between the legs of the ladder stops regardless of $\gamma$ value. It is worth mentioning that similar 
independence {of the} flux is also observed in the entanglement entropy for uncorrelated and ``spin disorder''. 

\section{ Participation ratio}
\label{secPR}

\begin{figure}
  \includegraphics[width=\linewidth]{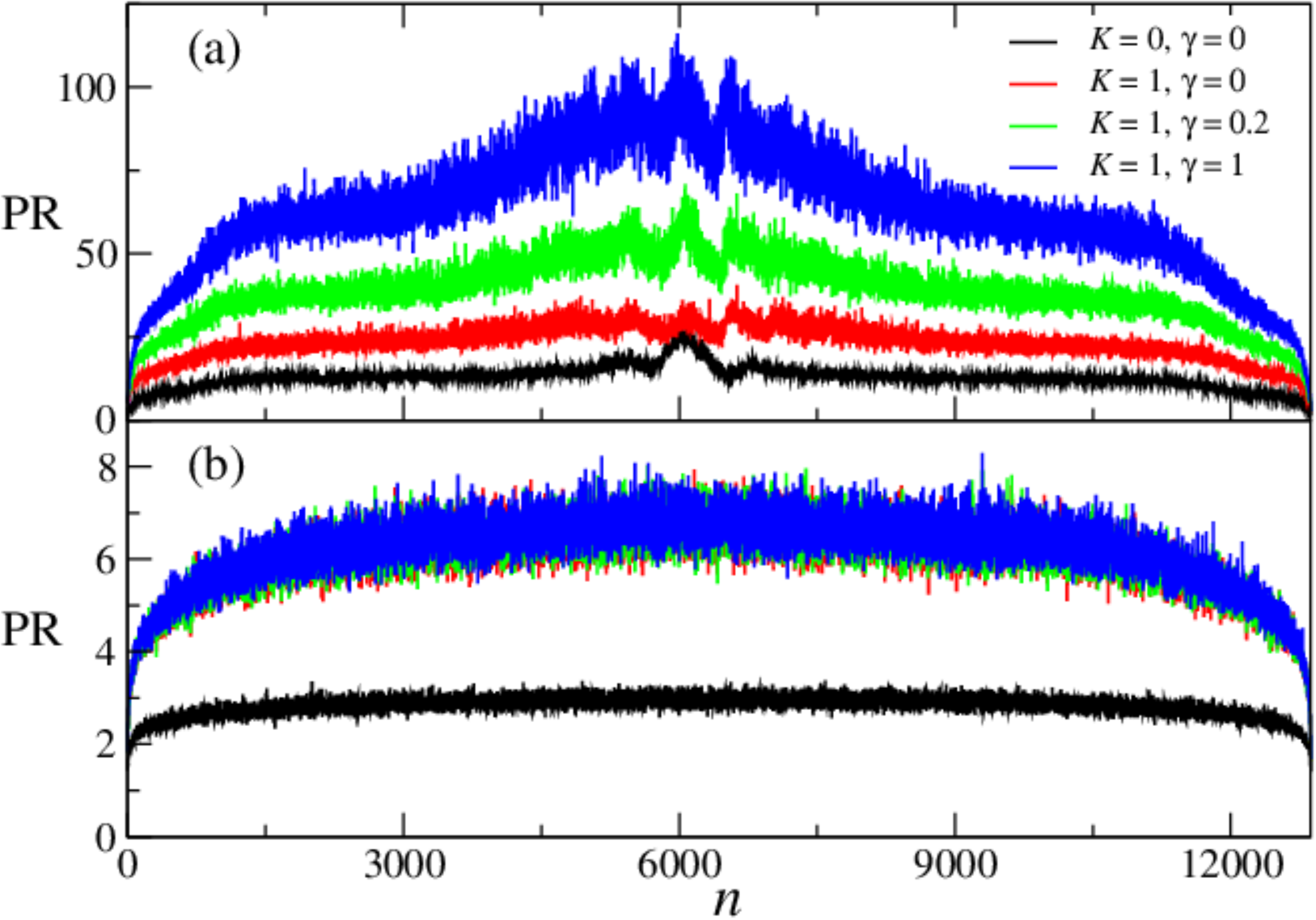}
  \caption{Averaged over disorder realizations participation ratio (a) for Hamiltonian \eqref{model} and (b) for disorder 
   	   uncorrelated between spin components - leading to full MBL of both charge and spin.}
\label{fig:ipr}
\end{figure}
One may make the observation {that the disorder averaged quantities do not depend on the} synthetic flux  even stronger. Consider the participation ratio of $n$-th eigenstate $|n\rangle$:
\begin{equation}
 {\cal PR}_n=1/\sum_i |\langle n|b_i\rangle|^4,
 \label{eq:PR}
\end{equation}
where $|b_i\rangle$ are basis functions (we chose a natural basis of Fock states on lattice sites). We denote as  $\mathrm{PR}_n$ the  participation ratio of level $n$ averaged over disorder realizations. For identical random 
disorder for both spin components [Hamiltonian \eqref{model}] PR defined in this way depends on $K$ and $\gamma$ - compare 
Fig.~\ref{fig:ipr}(a). Significant values of PR suggest that localization is not complete. For the disorder uncorrelated between spin 
components with the same amplitude $W$ we observe significantly smaller PR (pointing towards stronger localization), moreover, PR becomes 
independent of $\gamma$ (recall that PR$_n$ are obtained after disorder averaging). Thus not only $C(t)$ or $S(t)$ but also other 
observables may be expected to be $\gamma$-independent once the spin component is localized. 

In this respect we observe a clear asymmetry between a real dimension (along the chain) and the synthetic dimension (represented by spin 
components). The localization of the latter is essential for $\gamma$-independence of disorder averaged observables while the charge 
localization plays little role. Observe that for pure ``spin disorder'', as shown in Fig.~\ref{corr_spin} the correlations for $K=1$ are 
flux independent while charge degrees of freedom are delocalized.

\begin{figure}
  \includegraphics[width=\linewidth]{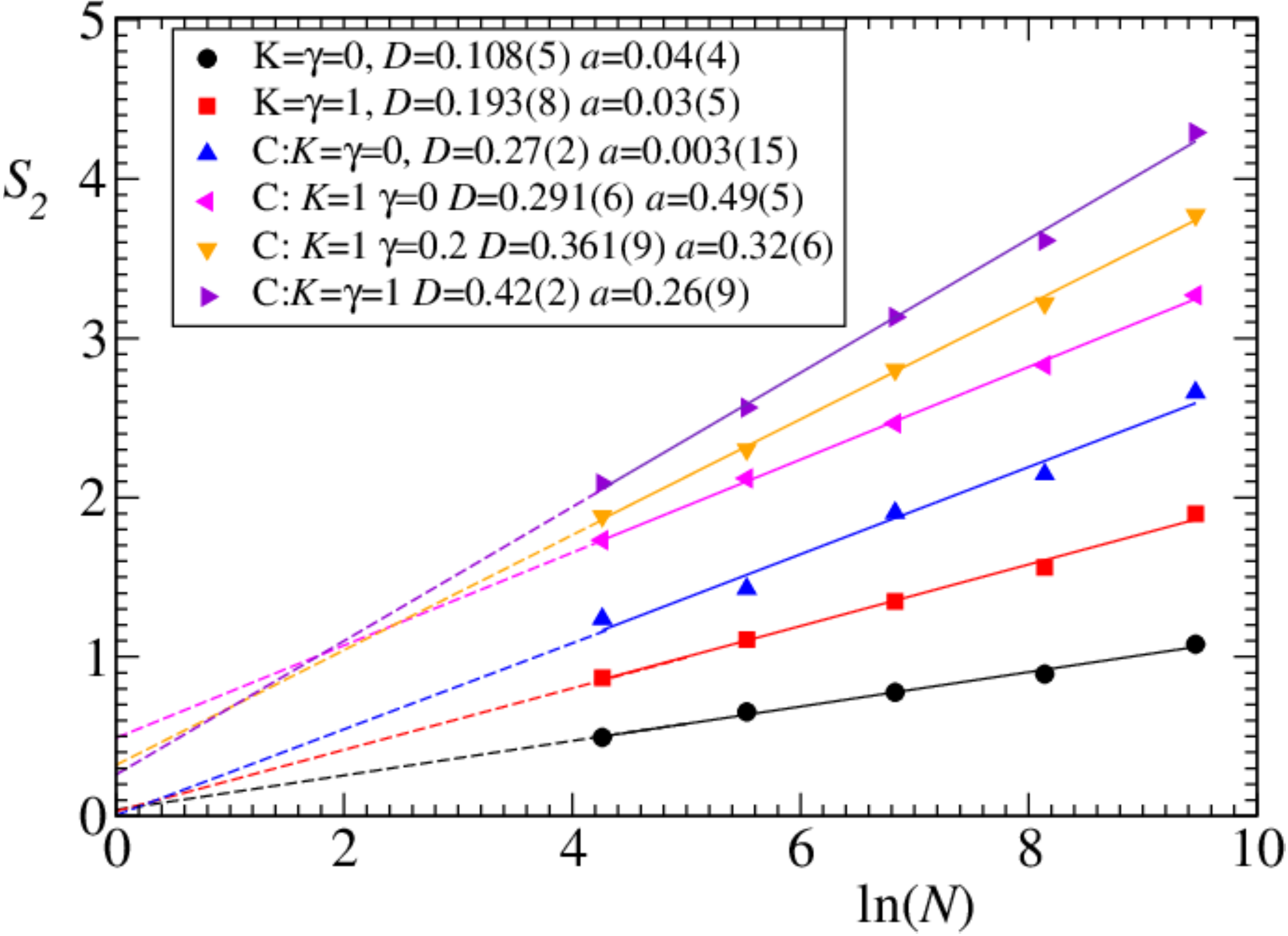}
  \caption{Participation entropy, {$\overline {S_2}$}, as a function of the {logarithm of} Hilbert space {dimension} for different cases considered. 
  {Two low lying lines show the linear dependence of $\overline {S_2}$ on $\ln(N)$ for independent disorder for up and down fermions, with TRI symmetry broken case given by 
  red squares. Other four cases corresponds to identical disorder for both legs (correlated disorder denoted by C in the legend). The leading slope $D$ in all cases is lower than 0.5
	indicating multifractal character of eigenvectors resembling the finding of \cite{Mace19}.} 
   }
\label{fig:ipr2}
\end{figure}
The conclusions drawn from participation ratio at a given system size may be qualitative only, it is important to study their
behavior as a function of the system size as described, e.g. in \cite{Mace19}. Following the prescription {of \cite{Mace19} we} 
define the participation entropy $S_2(n) = \ln {\cal PR}_n = -\ln \sum_i |\langle n|b_i\rangle|^4$ where the latter equality
gives a relation of $S_2$ with the inverse of the participation ratio \eqref{eq:PR},
commonly known as the inverse participation ratio (IPR). For  states in the middle of the spectrum
we define $\overline{S_2}$ as an average over both the different levels and the disorder realizations and refer to it as a
participation entropy. As shown in \cite{Mace19} one should consider how $\overline{S_2}$ behaves with the system size.
In particular for perfectly delocalized state $\overline{S_2}= \ln N$ where $N$ is the Hilbert space dimension while for
perfectly localized state  $\overline{S_2}$ should be a constant independent of $N$.
For spinless fermions, the paradigmatic MBL system it was found \cite{Mace19} that $\overline{S_2}$ is well fitted by 
$\overline{S_2}=D\ln N +a$ finding in particular that, on the localized side $D\approx 0.088$ and $a>0$ in the Fock basis we consider
(the obvious drawback of {the participation entropy} is that  {it depends on the choice of basis}). Together with other
measures and considering much larger {system} sizes that led the authors to conclude that eigenstates on the MBL side are not fully localized and reveal multifractal character. 

Following this example we looked at the scaling of $\overline{S_2}$ for different combinations of $K$ and $\gamma$ parameters as well as different disorder cases. Again we chose $W=32$ and system sizes up to {$L=8$}. The results are collected in Fig.~\ref{fig:ipr2}. As for spinless fermions, the obtained participation entropy values increase linearly with $\ln N$. For ``full MBL'' i.e. for independent disorder for up and down fermions the slope of the lines is smallest with $D=0.108(5)$ for a standard Hubbard case i.e. for $K=\gamma=0$.
The presence of a gauge field, $K=\gamma=1$ increases $D$ twice but the general trend remains  the same. For the same disorder for  up and down fermions (correlated disorder) 
the increase of  $\overline{S_2}$ with $\ln N$ is faster but follows the same trend with $0<D<0.5$ which   suggests  the multifractal character of eigenvectors.
We shall also note that the subleading free term $a$ in all cases studied is positive as observed for a localized side of the transition \cite{Mace19}.

The limited {system} sizes considered by us do not allow us to proceed further with a full finite size scaling analysis. However, the data presented in Fig.~\ref{fig:ipr2} strongly suggest that the multifractal character of wavefunctions in spinless fermionic system postulated by \cite{Mace19} similarly appears for spinfull fermions studied.

\section{Conclusions}
\label{sec:conc}

We have analyzed properties of the disordered chain of spin-1/2 fermions in the presence of the synthetic magnetic field. While the 
spectral properties such as the average gap ratio indicate the transition to many-body localized phase, the time dynamics 
suggest that MBL is realized in charge (density) sector only. The presence of the synthetic magnetic field delocalizes the spin sector 
as revealed by the decay of spin time correlation function. Similarly, the entropy of entanglement in the system grows much faster in 
the presence of the magnetic field flux. 

Interestingly the spectral properties of the system strongly depend on the realized symmetries providing a nice example of the effects 
due to a generalized TRI (i.e. a TRI combined with a discrete symmetry). In effect, despite the standard time reversal symmetry being 
broken the spectral properties in the delocalized regime resemble that of GOE unless additional symmetry breaking terms are introduced 
into the model.

A comparison of the dynamics of correlation functions for fermions and hard-core bosons in the absence of the flux ($\gamma=0$) but when the 
driving term $K$ couples up and down polarized particles shows sensitivity of the observed phenomena to quantum statistics. For fermions, due 
to their commutation relations, a kinetic energy part corresponding to the transition between spin up and down fermions commutes with the 
rest of the Hamiltonian. In effect, the exact quantum dynamics is given by a rapidly oscillating solution whose one envelope is given by 
$K=0$ (i.e. no Rabi coupling) solution. This behavior is absent for hard-core bosons.

Last but not least, we considered different types of random disorder, in particular spin dependent disorder that leads to strong 
localization in the spin sector. In such a case time-dependent observables become, after disorder average, independent of flux $\gamma$. 
Analysis of participation ratio suggests the multifractal character of eigenvectors in the strong disorder regime also in the presence of artificial flux.
We believe that the system sizes examined in present work are sufficient to grasp robust features of considered systems on time scales 
relevant for experiments with ultracold atoms.

\acknowledgments
 Interesting discussions with T. Chanda, D. Delande and K. \.Zyczkowski on subjects related to this work are acknowledged. 
The support by PL-Grid Infrastructure was important for this work. This research has been supported by National Science 
Centre (Poland) under projects 2015/19/B/ST2/01028 (P.S.), 2018/28/T/ST2/00401 (doctoral scholarship -- P.S.) and 
2016/21/B/ST2/01086 (K.S. and J.Z.).


%

\end{document}